\documentclass[]{emulateapj}

\usepackage{amsmath}
\usepackage{graphicx}
\usepackage{color}
\usepackage{rotating}
\usepackage{lscape}

\usepackage{ulem}

\shorttitle{
Dynamics of Non-Steady Spiral Arms in Disk Galaxies}
\shortauthors{Baba, Saitoh \& Wada}

\begin{document}

\title{Dynamics of Non-Steady Spiral Arms in Disk Galaxies}

\author{
	Junichi \textsc{Baba}\altaffilmark{1}, 
	Takayuki R. \textsc{Saitoh}\altaffilmark{1}, and 
	Keiichi \textsc{Wada}\altaffilmark{2}
}

\altaffiltext{1}{Interactive Research Center of Science,
Graduate School of Science and Engineering, Tokyo Institute of Technology
2--12--1 Ookayama, Meguro, Tokyo 152--8551}

\altaffiltext{2}{Graduate School of Science and Engineering, Kagoshima University,
    1--21--30 Korimoto, Kagoshima, Kagoshima 890--8580.}

\email{babajn@geo.titech.ac.jp}

\begin{abstract}
In order to understand the physical mechanisms underlying non-steady
stellar spiral arms in disk galaxies, we analyzed the growing and damping
phases of their spiral arms using three-dimensional $N$-body simulations.  We
confirmed that the spiral arms are formed due to a swing amplification
mechanism that reinforces density enhancement as a seeded wake.  In
the damping phase, the Coriolis force exerted on a portion of the arm
surpasses the gravitational force that acts to shrink the portion. 
Consequently, the stars in the portion escape from the arm, and
subsequently they form a new arm at a different location.  The
time-dependent nature of the spiral arms are originated in the continual
repetition of this non-linear phenomenon.  Since a spiral arm does not rigidly rotate, 
but follows the galactic differential rotation, the stars in the
arm rotate at almost the same rate as the arm. In other words, every
single position in the arm can be regarded as the co-rotation point.
Due to interaction with their host arms, the energy and angular momentum
of the stars change, thereby causing the radial migration of the stars.
During this process, the kinetic energy of random motion (random energy)
of the stars does not significantly increase, and the disk remains dynamically cold.
Owing to this low degree of disk heating, the short-lived spiral arms can
recurrently develop over many rotational periods.  The resultant
structure of the spiral arms in the $N$-body simulations is consistent
with some observational nature of spiral galaxies.  We conclude that
the formation and structure of spiral arms in isolated disk galaxies
can be reasonably understood by non-linear interactions between a
spiral arm and its constituent stars.
\end{abstract}


\section{Introduction}

The spiral arms of disk galaxies are the most prominent structures, 
and the arms are formed due to gravitationally driven variations in 
the surface density in the stellar disk 
\citep{RixZaritsky1995,Grosbol+2004,Zibetti+2009,Elmegreen+2011}. 
It is known that the spiral arms of disk galaxies can be excited by 
tidal interactions with nearby companion galaxies 
\citep[e.g.,][]{Oh+2008,Dobbs+2010,Struck+2011}, 
as well as by the central stellar bar \citep[e.g.,][]{SellwoodSparke1988}.
In addition, spiral arms can also be self-induced and maintained without 
these external gravitational perturbations, 
and further, they can propagate stationary density waves in globally stable disks, 
as hypothesized by \citet{LinShu1964}.
However, the physical origin and dynamical evolution of spiral arms
in disk galaxies have thus far not been fully understood.
In this study, we focus on the dynamics of stellar spiral arms and 
stars around them in a disk galaxy
without external perturbations and a bar structure.

The  Lin-Shu theory posits that the stellar spiral arms can be  
interpreted as (quasi-) stationary density waves with a constant pattern speed 
\citep{LinShu1964,BertinLin1996}; 
the theory has gained wide acceptance as providing a clear explanation regarding 
the dynamics of spiral arms. 
However, as pointed out in the study by \citet{Toomre1969}, 
the quasi-stationary hypothesis has a serious limitation 
because of a dispersal nature of the tight winding spiral waves, 
they radially propagate with the group velocity, 
and are consequently absorbed at the Lindblad resonances
\citep{LyndenBellKalnajs1972}.
Therefore, the stationary waves that last for a long duration require
some amplification mechanisms such as WASER \citep{Mark1976}
and a feedback cycle that involves the reflection of the inward propagating 
wave into an outward propagating one at the $Q$-barrier \citep{Bertin+1989a,Bertin+1989b}.

However, nearly all the previous time-dependent simulations 
that have been executed thus far have been unable to prove 
the existence of stationary density waves in 
a disk galaxy without external perturbations and a bar structure
\footnote{
\citet{Thomasson+1990} and \citet{ElmegreenThomasson1993}
have suggested that the feedback cycle caused by the reflection at the $Q$-barrier
causes the long life of the $m=2-3$ spiral arms. 
However, their $N$-body spirals are not stationary; instead, a time-dependent evolution 
of the spiral features is observed. See also \citet{Sellwood2011}.
}.
\citet{SellwoodCarlberg1984} claimed that self-induced stellar spiral structures 
in stellar disks are not stationary and fade away within an interval of about ten rotational periods.
Their study stresses on the significance of cooling mechanisms, such as dissipation 
by the interstellar medium (ISM), that are required to maintain stellar spiral arms.
More recently, \citet{Fujii+2011} used 3-D $N$-body simulations to show
that although stellar spiral arms are short-lived, they are also recurrently formed, and 
as a result, the spiral features are maintained over 10 Gyr \citep[see also][]{Sellwood2011}. 
This is because of negative feedback that causes the dynamical heating of stars 
in the disk galaxies due to spirals.
The non-steady spirals are also subject to a similar phenomenon 
when the dynamics of the ISM are self-consistently solved with stars in disk galaxies.
Both the stellar spirals and the ISM undergo motion in the wake of 
the the galactic rotation \citep[see also][]{Grand+2012}.
In fact, the spirals can be considered as being ``wound''.
The ISM forms dense regions associated with the non-steady 
stellar spirals; however, these regions are not the conventional ``galactic shocks''\citep{Wada+2011}.

The existence of non-steady stellar spirals has been suggested by certain observations in our Galaxy.
For example, the age-velocity dispersion relation of stars in the solar neighborhood 
\citep[e.g.,][]{Holmberg+2007} most naturally accounts for the existence of the non-steady spirals 
\citep{CarlbergSellwood1985,BinneyLacey1988,JenkinsBinney1990,deSimone+2004}.
\citet{Baba+2009} analyzed the kinematics of star-forming regions
using numerical simulation data of a barred galaxy such as the Milky Way galaxy \citep{Baba+2010},
and they found that the kinematics of these regions are consistent with the observed peculiar 
motions of maser sources. These studies support the view that the Galactic stellar spiral arms 
are non-steady instead of stationary density waves.

However, the current theoretical understanding of the dynamics of 
non-steady spirals is thus far insufficient.
It has been suggested that a key factor towards understanding their dynamics 
is the orbital evolution of stars. In this study, we analyze the orbits of stars 
associated with non-steady spirals in the growth and damping phases.

In Section \ref{sec:Method}, we summarize the numerical model and the method,
which are the same as reported by \citet{Wada+2011}, 
except that the present model does not include the ISM, i.e., 
our study involves pure $N$-body simulations.
We examine the global evolution of the stellar disk in Section \ref{sec:GlobalMorphology},
in which we show the non-steady nature of spirals developed in the galactic disk.
In Section \ref{sec:SpiralDynamics}, we focus on a typical spiral arm and describe its dynamical evolution, 
i.e., the amplification (Section \ref{sec:AmplifySpiralArm}) and 
destruction (Section \ref{sec:DampingSpiralArm}) processes.
Section \ref{sec:StarAroundSpiralArm} examines the evolution of stars around the growing spiral arm 
from the viewpoint of the angular momentum-energy space.
Finally, in Section \ref{sec:Discussion}, we compare our simulation results with observations, 
and we provide a perspective view on grand-design spirals. 
The grand-design spirals (i.e.,  $m=2$ spirals)
\footnote{
Although the term grand-design spiral indicates a reasonably coherent and extensive spiral arm
in the stellar mass distribution \citep[See][for a more precise definition]{ElmegreenElmegreen1982},
in a majority of cases, this means that the galaxy has $m=2$ spiral arms \citep{Kendall+2011}.
} 
will be the subject of our future studies.

\section{Models and Methods}
\label{sec:Method}

The simulations of the formation and evolution of spiral arms 
were carried out using a three-dimensional $N$-body simulation.
The initial point of our simulation was a disk galaxy in a nearly equilibrium state.  
The density profile of the stellar disk in this case is given by 
\begin{equation}
\rho_{\rm sd}(R,z) = \frac{M_{\rm sd}}{4\pi {R_{\rm sd}}^2 z_{\rm sd}}
\exp(-R/R_{\rm sd}) {\rm sech}^2(z/z_{\rm sd}),\label{eq:stellardisk}
\end{equation}
where $M_{\rm sd}$ is the mass of the stellar disk, $R_{\rm sd}$ is the
radial scale-length, and $z_{\rm sd}$ is the vertical scale-length. 
Since the distribution function of a stellar disk is unknown, 
we generated the equilibrium state of the stellar disk 
by using an empirical method: We first generated a stellar
disk in a near-equilibrium state based on a Maxwellian approximation
\citep{Hernquist1993}, and we subsequently allowed the disk 
to evolve for 6 Gyr under the constraint of axisymmetry
\citep{McMillanDehnen2007,Fujii+2011}. 

\begin{figure*}[htbp]
\begin{center}
\includegraphics[width=.9\textwidth]{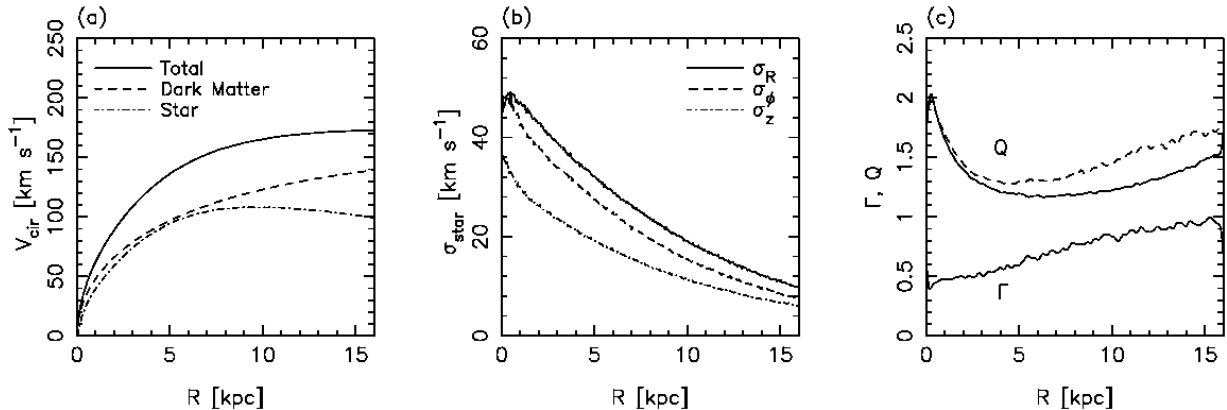}
\caption{
	(a) Initial circular velocity curves 
	(solid line: circular rotation, dashed: line dark matter, dot-dashed line: stars). 
	(b) Stellar velocity dispersions 
	(solid line: $\sigma_{\rm R}$, dashed line: $\sigma_{\phi}$, dot-dashed line: $\sigma_{\rm z}$).
	(c) Initial $Q$- and $\Gamma$-profiles. 
	The term $\Gamma$ denotes a dimensionless shear rate of the disk as defined by 
	the expression $-d\ln\Omega/d\ln R$, where $\Omega$ denotes the angular frequency.
	The dashed curve presents the final $Q$-profile ($T_{\rm rot}= 15$).
	}
\label{fig:rotcurve}
\end{center}
\end{figure*}

We modeled a fixed potential of the dark matter (DM) halo 
whose density profile follows the NFW profile 
\citep{Navarro+1997}:
\begin{equation}
\rho_{\rm h}(r)=\frac{\rho_0}{r/r_{\rm s}(1+r/r_{\rm s})^2},
\end{equation}
where
\begin{equation}
\rho_0 = \frac{M_{\rm h}}{4\pi {R_{\rm h}}^3}
\frac{{C_{\rm NFW}}^3}{\ln(1+C_{\rm NFW})+C_{\rm NFW}/(1+C_{\rm NFW})},
\end{equation}
\begin{equation}
  C_{\rm NFW} = R_{\rm h}/r_{\rm s},
\end{equation}
an the terms $C_{\rm NFW}$, $M_{\rm h}$, and $R_{\rm h}$ denote the halo concentration
parameter of the halo profile, the DM halo mass, and the virial radius, respectively.  
The model parameters adopted here are summarized in Table \ref{tbl:models}.
The circular velocity, velocity dispersions, $Q$-values and $\Gamma$-profile 
of the stellar disk are shown in Figure \ref{fig:rotcurve}.
Here, $\Gamma$ denotes a dimensionless shear rate of the disk as defined by 
the expression $-d\ln\Omega/d\ln R$, where $\Omega$ denotes the angular frequency.

We used the simulation code {\tt ASURA} \citep{Saitoh+2008,Saitoh+2009}, where
the self-gravity of stellar particles is calculated by the Tree with 
the GRAPE method \citep{Makino1991}.  
We used a software emulator of GRAPE, the Phantom-GRAPE \citep{Tanikawa+2012}.  
The opening angle was set to $0.5$
with the center-of-mass approximation.  To perform a time integration, we used a
leapfrog integrator with variable and individual time-steps.  We used a
Plummer softening length of $\epsilon = 30$ pc.  This value is sufficiently small
to resolve the three-dimensional structure of a disk galaxy
\citep{Hernquist1987,Bottema2003}.  It is to be noted that the formation of the spiral arms is
inhibited if the adopted softening length is considerably larger than the disk thickness
($\sim 300$ pc). The number of particles is 300 million.

\begin{table}[htdp]
\caption{Model parameters for each mass component in $N$-body simulation}
\begin{center}
\begin{tabular}{ lll }
\hline
\hline 
 Component	&	Parameter	& Value	\\
\hline
Dark halo			& Mass ($M_{\rm h}$)		&	$6.3\times 10^{11}$ M$_{\odot}$\\
(Rigid)			& Radius ($R_{\rm h}$)		&  $122$ kpc\\
				& Concentration ($C_{\rm NFW}$) 	& $5.0$ \\
\hline
Initial stellar disk	& Mass ($M_{\rm sd}$)		&  $3.2\times 10^{10}$ M$_{\odot}$\\
(Live)			& Scale length ($R_{\rm sd}$)	& $4.3$ kpc \\
				& Scale height	($z_{\rm sd}$)	& $0.3$ kpc\\
\hline
\end{tabular}
\end{center}
\label{tbl:models}
\end{table}%

\section{Global Evolution of Spirals}
\label{sec:GlobalMorphology}

\begin{figure*}
\begin{center}
\includegraphics[width=.9\textwidth]{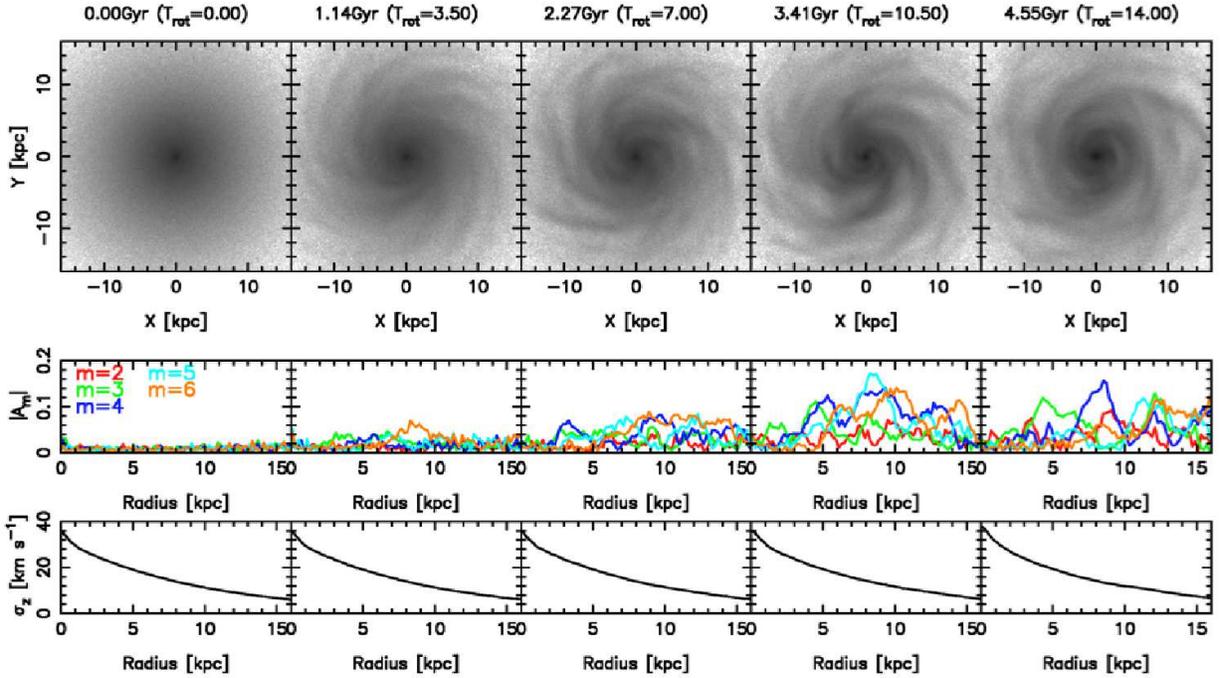}
\caption{
	(Top) Evolution of stellar disk. 
	The disk's surface density is shown in the logarithmic scale.
	The rotation direction of the disk is counterclockwise.
	Here, a rotational period, $T_{\rm rot}$, is measured at $R = 2 R_{\rm sd}$, 
	and corresponds to 325 Myr.
	(Middle) The radial distributions of the spiral modes analyzed by 
	the Fourier decomposition (see text).
	(Bottom) The radial distributions of the vertical velocity dispersion ($\sigma_z$)
	of the stellar disk.
}
\label{fig:snapshotsMS}
\end{center}
\end{figure*}

Figure \ref{fig:snapshotsMS} shows the time evolution of the stellar disk.
Hereafter, we use a rotational period measured at $R = 2 R_{\rm sd} (=8.6~\rm kpc)$, $T_{\rm rot}$, 
which corresponds to 325 Myr. The top panels show the face-on views of the stellar disk.
The middle panels show the radial distributions of 
the spiral modes which are analyzed by performing a one-dimensional 
Fourier decomposition of the disk surface density using the polar coordinates $(R,\phi)$:
\begin{eqnarray}
\frac{\Sigma(R,\phi)}{\Sigma_{\rm 0}(R)} 
	=  \sum_{m=0}^{\infty} A_m(R)\cos\{m[\phi-\phi_m(R)]\}.
\end{eqnarray}
Here, $A_m$ and $\phi_m(R)$ denote the Fourier amplitude and phase angle 
for the $m$-th mode, respectively \citep[e.g,][]{RixZaritsky1995}.
Upon execution of the simulation, spiral patterns initially developed from the noise 
with an $e$-folding time of about 4 galactic rotations ($T_{\rm rot} \sim 4$), 
and subsequently, the patterns settled to a nearly constant level, 
consistent with the report by \citet{Fujii+2011}.
While the global features were in the quasi-steady state, 
the local spiral features were not static.
The most prominent mode always changed on a rotational time scale, and
it also showed radial dependence.

\begin{figure*}
\begin{center}
\includegraphics[width=.9\textwidth]{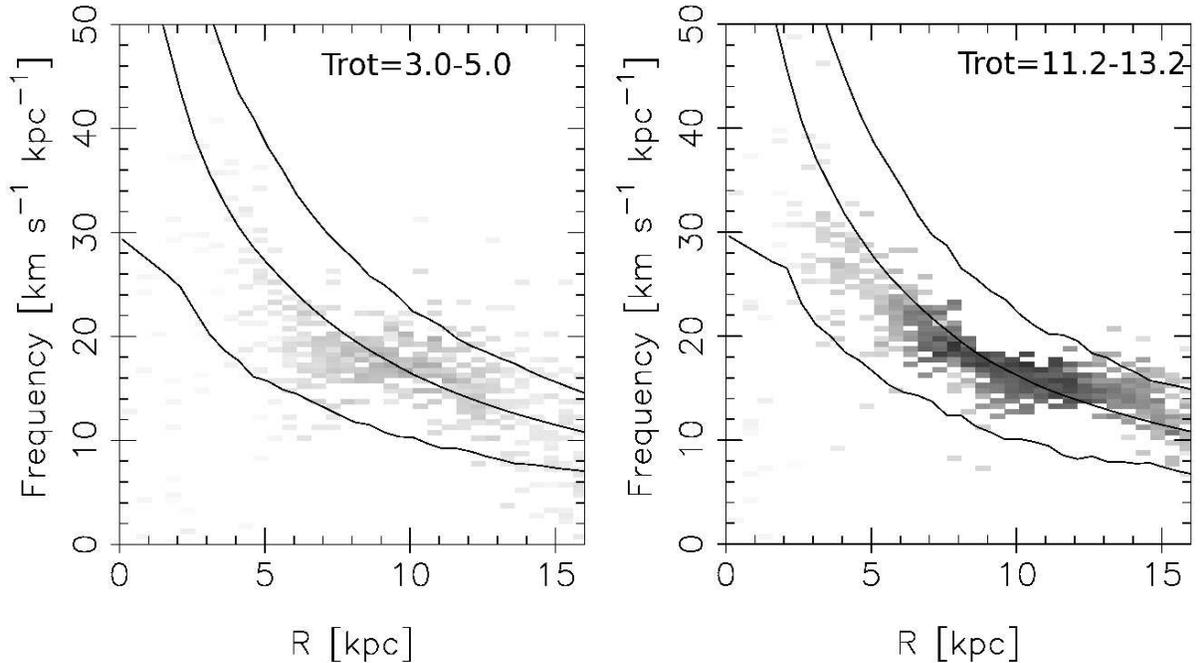}
\caption{
	Pattern speeds of dominant mode ($m=4$) for $T_{\rm rot} \simeq 4.0$ (left)  and $12.2$ (right).
	The pattern speed is analyzed by measuring the rate of change of phase $\phi_{m=4}$
	at each annulus every 10 Myr, and these phases are combined over the periods 
	$T_{\rm rot} = 3.0-4.0$ and $11.2-13.2$.
	The contours show the amplitude, $|A_{m=4}(R)|$, of the dominant mode.
	The superimposed curves show the radial variations in $\Omega$, $\Omega \pm \kappa/4$.
}
\label{fig:PatternSpeed}
\end{center}
\end{figure*}

An important dynamical feature of the time-dependent multi-arm spirals is
that the spirals ``co-rotate'' with the galactic rotation. 
Figure \ref{fig:PatternSpeed} shows 
the angular pattern speed of a dominant stellar spiral ($m=4$).
The angular frequency is analyzed by the rate of change of phase 
$\phi_{\rm m=4}(R)$ during two galactic rotation periods.
The pattern speed is not constant. 
In fact, the pattern speed decreases with radius in a manner similar to galactic rotation for any radii.
This result shows that there is no ``single'' co-rotating point in the disk,
in contrast to stationary spiral waves (or bar) with a rigid rotation.
Therefore, no single spiral can last for more than one rotational period 
(see Section \ref{sec:StarAroundSpiralArm}).
\citet{Wada+2011} and \citet{Grand+2012} have also reported that the spiral arms in 
their simulations are co-rotating, winding and short-lived.
Observationally, the phenomenon of co-rotating spiral arms has been confirmed in at least 
a few of the nearby spiral galaxies such as M51, NGC 1068, M101, IC 342, NGC 3938, and NGC 3344
using the Tremaine-Weinberg (TWR) analysis \citep{Merrifield+2006,Meidt+2008,Meidt+2009}.

\begin{figure*}
\begin{center}
\includegraphics[width=.9\textwidth]{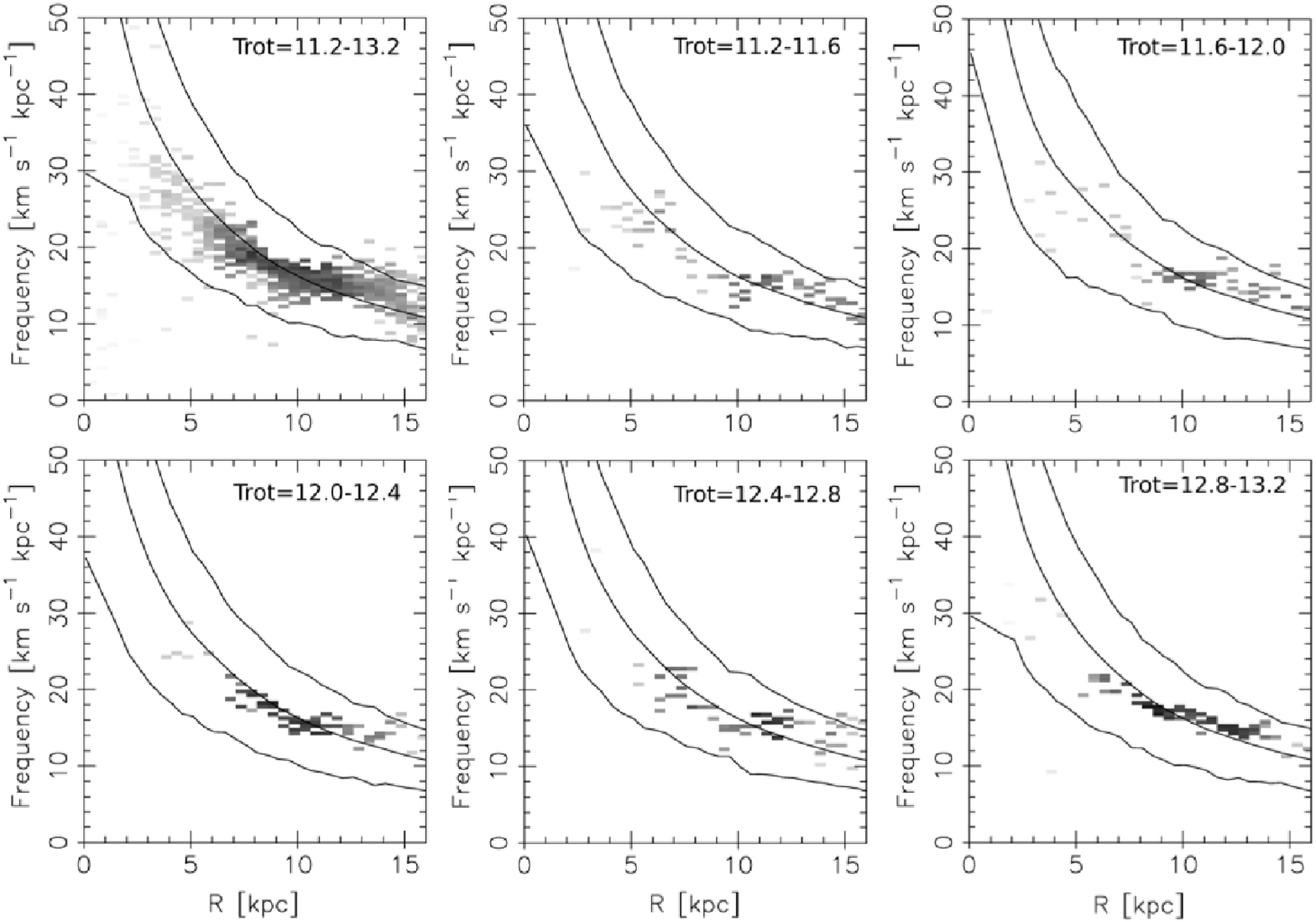}
\caption{
	Pattern speeds of dominant mode ($m=4$) for $T_{\rm rot} \sim 12$.
	The pattern speed is analyzed by measuring the rate of change of phase $\phi_{m=4}$
	at each annulus every 10 Myr during the periods $T_{\rm rot} = 11.2-13.2$, $11.2-11.6$, $11.6-12.0$, 
	$12.0-12.4$, $12.4-12.8$, and $12.8-13.2$.
	The contours show the amplitude, $|A_{m=4}(R)|$, of the dominant mode.
	The superimposed curves show the radial variations in $\Omega$, $\Omega \pm \kappa/4$.
}
\label{fig:PatternSpeedAppendix}
\end{center}
\end{figure*}

It is noteworthy that the pattern speed curve seems slightly flatter 
(particularly when $R \sim 10-15$ kpc) than the circular rotation curve.
In order to investigate the cause of the flattening, we divided this period into five periods and 
performed the same analysis for each period (Figure \ref{fig:PatternSpeedAppendix}).
The pattern speeds in the all periods decrease with increasing radius in a manner similar to 
galactic rotation for any radii; however, they show slightly flatter distributions at  certain instants 
(e.g., $T_{\rm rot}= 12.4-12.8$).  
This flatter distribution lasts for less than one galactic rotation period ($\Delta T_{\rm rot}<1$).
Thus, this structure is not a long-lasting structure.

Further, it is significant that the gravitational scattering of stars by
the spiral arms is not sufficiently large to erase all the spiral features.
The curves in Figure \ref{fig:rotcurve}c indicate that 
the value of $Q$ increases from an initial value of $1.2$ to a final value of $1.4$
around $R \approx 2 R_{\rm sd}$ at $T_{\rm rot} = 15$.
However, the spirals are not completely erased due to dynamical heating 
\citep[see][]{Fujii+2011}.
It is to be noted that heating effect along the vertical direction to the disk plane
is negligibly small (bottom panels in Figure \ref{fig:snapshotsMS}), 
which is consistent with the results of previous studies 
\citep[e.g.,][]{BinneyLacey1988,JenkinsBinney1990}.

\section{Growth and Damping Phases of Stellar Spiral Arms}
\label{sec:SpiralDynamics}

\begin{figure*}
\begin{center}
\includegraphics[width=.8\textwidth]{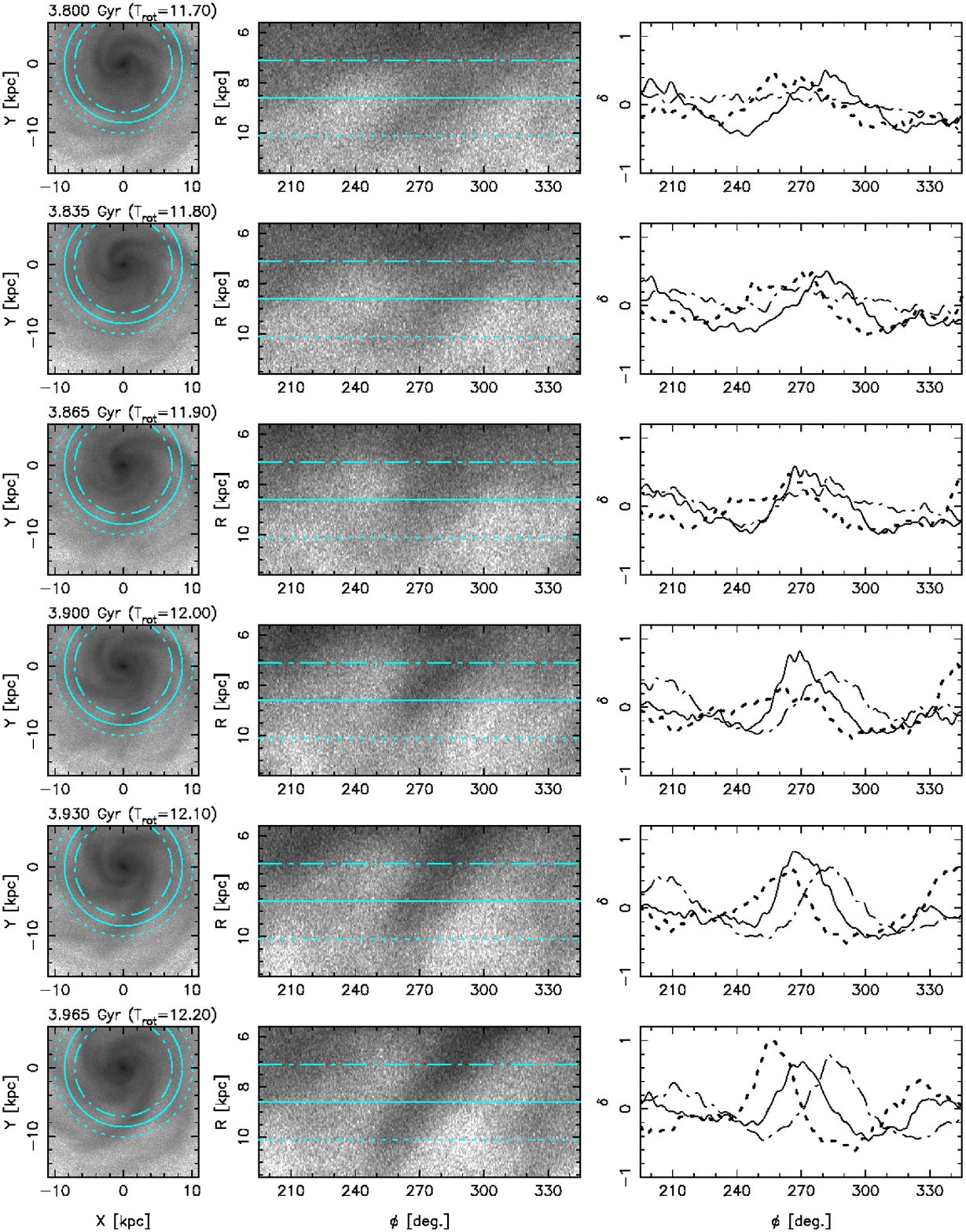}
\caption{
	Evolution of spiral arm in growth phase.
	Left columns: face-on views of the disk. 
	The surface density is shown in the logarithmic scale.
	The time interval for each phase of evolution is indicated in the upper corner of each panel to the left.
	The rotating frame at $R=2R_{\rm sd} = 8.6$ kpc (indicated by a solid circle) is adopted as a reference.
	The dotted and dot-dashed circles 
	indicate radii values of $R = 2R_{\rm sd}-1.5~\rm kpc$ and $2R_{\rm sd}+1.5~\rm kpc$, respectively.
	Middle columns: the density distribution corresponding to the spiral evolution along the $\phi-R$ plane.
	Right columns: the corresponding azimuthal density contrast ($\delta$) profiles 
	at $R = 2R_{\rm sd}-1.5~\rm kpc$ (dot-dashed), 
	$2R_{\rm sd}$ (solid), and $2R_{\rm sd}+1.5~\rm kpc$ (dotted).
}
\label{fig:snapshotModelMSgrowth}
\end{center}
\end{figure*}

\begin{figure*}
\begin{center}
\includegraphics[width=.8\textwidth]{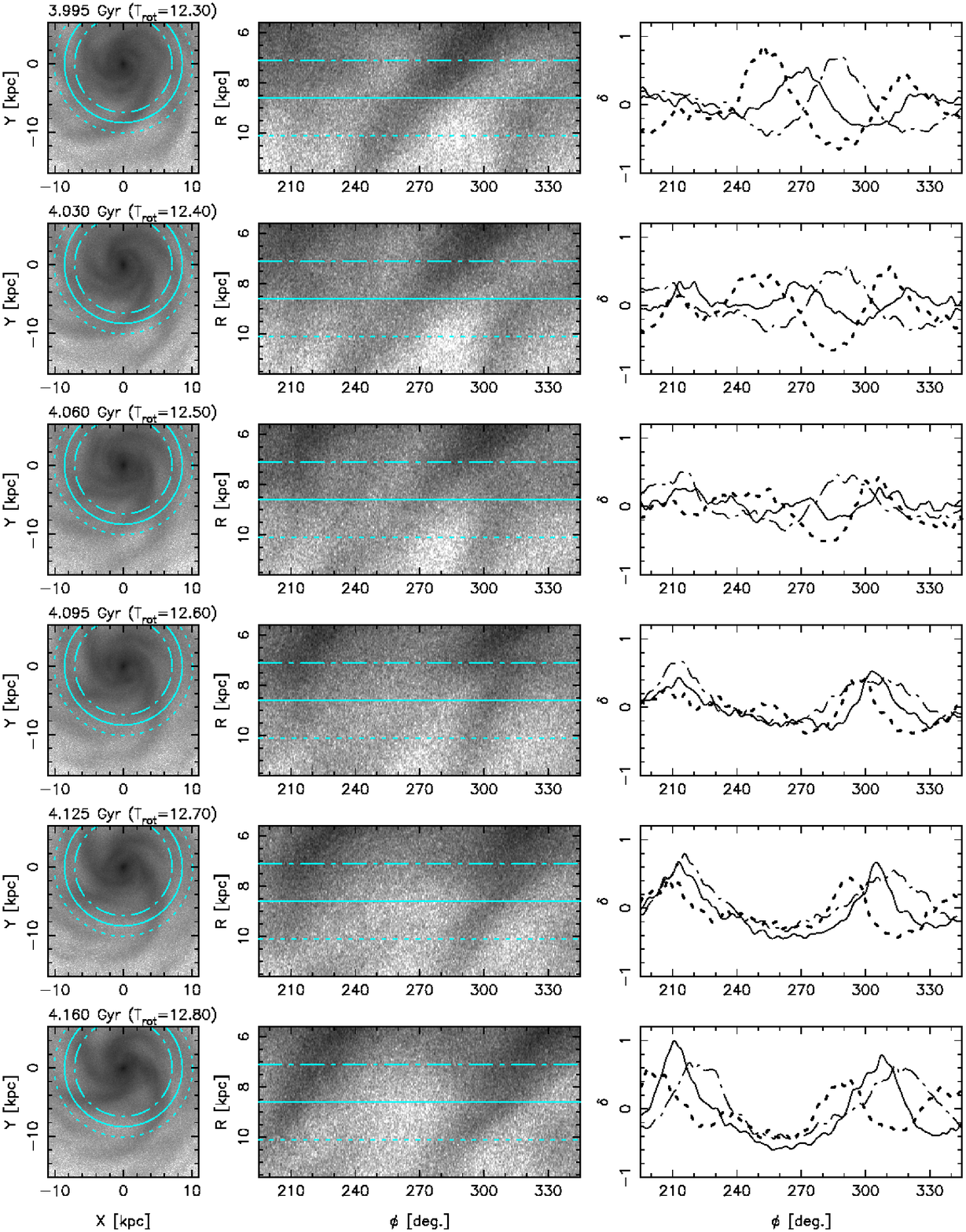}
\caption{
	Evolution showing thinning out of spiral ($T_{\rm rot} = 12.2-12.5$),  
	with subsequent reconnection with neighboring spiral arm ($T_{\rm rot} = 12.5-12.7$).
}
\label{fig:snapshotModelMSdamping}
\end{center}
\end{figure*}

In this section, we investigate the dynamical evolution of co-rotating spiral arms in detail. 
Figures \ref{fig:snapshotModelMSgrowth} and \ref{fig:snapshotModelMSdamping} show the
time evolution of a spiral arm in the co-rotating frame. 
A weak density enhancement can be observed around $\phi \simeq 240-300^{\circ}$ 
and $R \simeq 7-10$ kpc  for $T_{\rm rot} = 11.7$.
This density enhancement causes the growth of a prominent spiral arm until $T_{\rm rot} = 12.0$, 
and the arm has a maximum amplitude around $T_{\rm rot} \sim 12.2-12.3$. 
Beyond $T_{\rm rot} \simeq 12.3$, 
this spiral arm rapidly fades out ($T_{\rm rot} = 12.3-12.5$).
The arm merges with with a neighboring weak spiral arm ($T_{\rm rot} = 12.5-12.7$), 
and the peak density contrast again becomes $\delta \sim 1$ when $T_{\rm rot} = 12.8$.
In the following sections, we discuss the growth phase (Figure \ref{fig:snapshotModelMSgrowth})
and the damping phase (Figure \ref{fig:snapshotModelMSdamping} ) of the spiral separately.

\subsection{Growth Phase}
\label{sec:AmplifySpiralArm}

\begin{figure}
\begin{center}
\includegraphics[width=.45\textwidth]{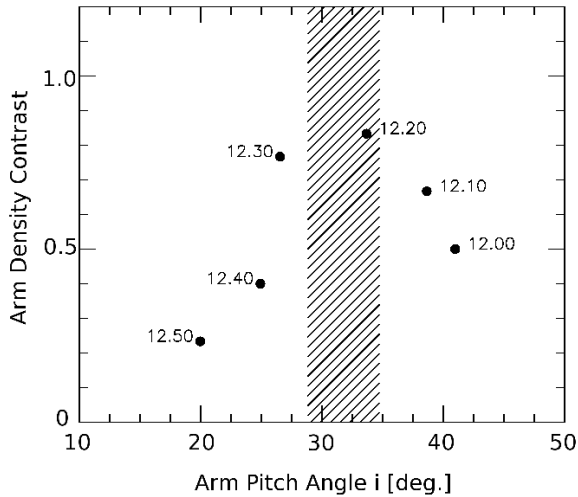}
\caption{
	Evolution of  spiral arm on $i-\bar{\delta}$ plane for $T_{\rm rot} = 12.0 - 12.5$. 
	The hatched region corresponds to the predicted maximum pitch angle 
	around the analyzed region ($Q \simeq 1.4$ and $\Gamma \simeq 0.75-0.85$) 
	due to swing amplification (refer to equation (7)).  
}
\label{fig:PitchAngleAmplitude}
\end{center}
\end{figure}

In the growth phase of the spiral arm ($T_{\rm rot} \simeq 12.0-12.2$), 
the pitch angle of the arm reduces due to the differential rotation of the disk.
This is clearly observed in the right hand panels of 
Figures \ref{fig:snapshotModelMSgrowth} and \ref{fig:snapshotModelMSdamping}; 
the two peaks at $2R_{\rm sd}+1.5$ kpc and $2R_{\rm sd}-1.5$ kpc
show increasing separation in terms of  $\phi$.
More quantitatively, Figure \ref{fig:PitchAngleAmplitude} shows the evolution of the spiral arm 
along the pitch angle - density contrast ($i-\bar{\delta}$) plane.
Here, we calculated the pitch angle ($i$) of the spiral arms using the relation 
\begin{eqnarray}
 \tan i = \frac{1}{R}\frac{\Delta R}{\Delta\phi}, 
\end{eqnarray}
where $R=2 R_{\rm sd}$ (= 8.6 kpc), $\Delta R = 3$ kpc, and $\Delta \phi$ is the azimuthal angle difference
between the contrast peaks at $R=2 R_{\rm sd} - 1.5$ kpc (dot-dashed lines) 
and $R=2 R_{\rm sd} + 1.5$ kpc (dotted lines) shown 
in Figures \ref{fig:snapshotModelMSgrowth} and \ref{fig:snapshotModelMSdamping}.
Here, we evaluated the positions of the contrast peaks at each radius by visual inspection.
The arm density contrast, $\bar{\delta}$, is calculated by averaging the corresponding density contrast
over the radial range in Figures \ref{fig:snapshotModelMSgrowth} and \ref{fig:snapshotModelMSdamping}.
As the pitch angle of the spiral arm decreases 
from $i \simeq 40^\circ$ ($T_{\rm rot} =12.0$) to $i \simeq 32^\circ$ ($T_{\rm rot} =12.20$),
the density contrast increases to a maximum, and subsequently, 
it decreases with increase in the pitch angle.
Thus, the spiral arm has a maximum amplitude when $i \sim 32^\circ$.

This amplification process associated with the galactic shear motion is known as 
the swing amplification \citep{Toomre1981,GoldreichLynden-Bell1965,JulianToomre1966,ToomreKalnajs1991}
\footnote{
\citet{DOnghia+2012} have investigated the growth of spiral arms via swing-amplification, 
and their nonlinear evolution is not fully consistent with 
the classic swing-amplification picture proved by \citet{JulianToomre1966}. 
}.
\citet{Fuchs2001a} calculated grid models of swing amplification process by 
varying the shear rate $\Gamma$ for $Q=1.4$. 
In his work, he derived a fitting empirical equation 
for swing amplification (eq.98 in his paper) as the following: 
\begin{eqnarray}
  \tan i_{\rm th} 
	= 1.932 - 5.186 \left(\frac{1}{2}\Gamma \right) + 4.704 \left(\frac{1}{2}\Gamma \right)^2,
\end{eqnarray}
where $i_{\rm th}$ is the pitch angle at which the spiral arm reaches the maximum amplitude.
We adopt this equation to evaluate the predicted pitch angle $i_{\rm th}$.
In our model, $\Gamma \simeq 0.8$ around $R = 2R_{\rm sd}$ (Figure \ref{fig:rotcurve}c).
Thus, substituting $\Gamma \simeq 0.8$ in eq. (7), we obtain $i_{\rm th} \simeq 32^\circ$.
This value is consistent with the evolution of the spiral arm with the maximum contrast 
(hatched region in Figure \ref{fig:PitchAngleAmplitude}).

\subsection{Damping Phase}
\label{sec:DampingSpiralArm}

\begin{figure*}
\begin{center}
\includegraphics[width=.90\textwidth]{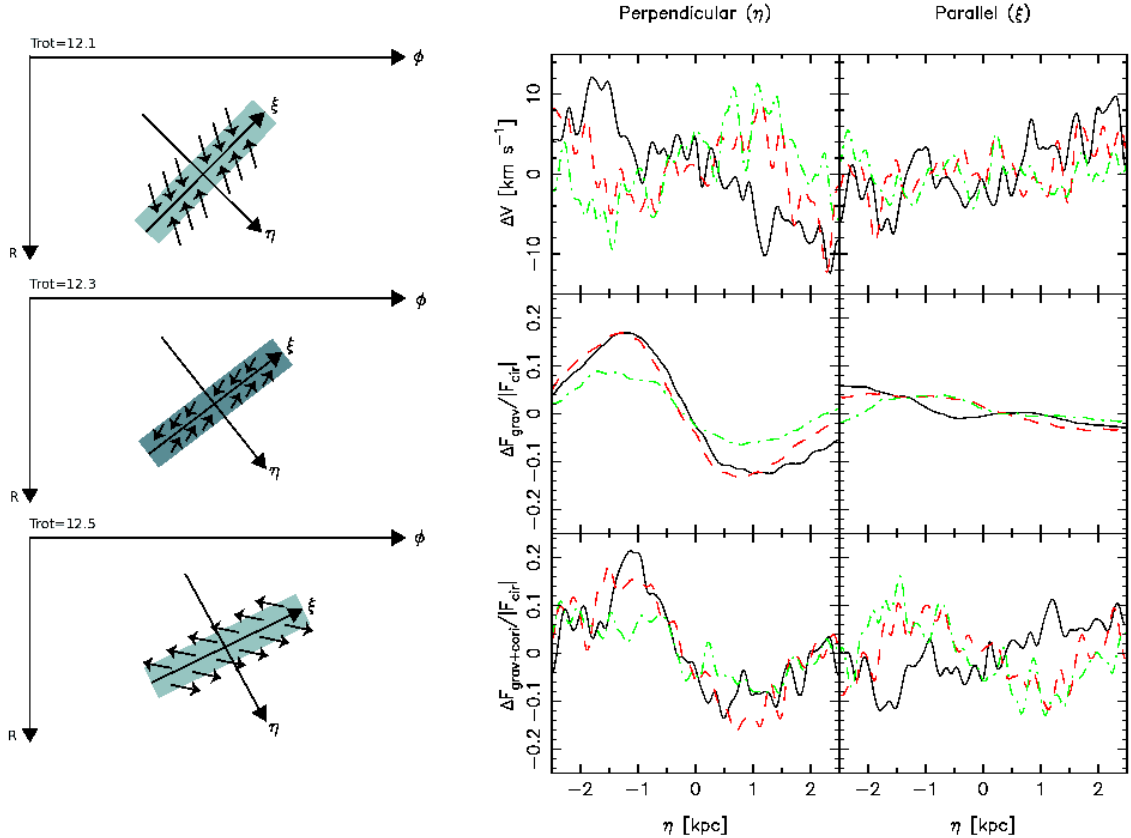}
\caption{
	(Left) Schematic view of velocity field with respect to non-steady spiral arm.
	The spiral arm is represented by the gray region.
	(Right) Top row: time evolution of relative velocities with respect to the stellar spiral arm ($\sim 270^\circ$),
	$\Delta v_\eta$ (left panel), $\Delta v_\xi$ (right panel), 
	at $T_{\rm rot}= 12.1$ (solid line), $12.3$ (dashed line), and $12.5$ (dot-dashed line). 
	The analysis is performed for the strip along the $\eta$-axis with a width of $|\xi|=500$ pc.
	Middle row: time evolution of non-axisymmetric gravitational forces (i.e., spiral perturbation)
	$\Delta F_{\rm grav} \equiv F_{\rm grav}-F_{\rm cir}$. 
	Here, $F_{\rm cir}$ denotes the axisymmetric gravitational force calculated by 
	the azimuthally averaged gravitational forces.
	Bottom row: time evolution of ``net'' non-axisymmetric forces 
	$\Delta F_{\rm grav+cori} (= \Delta F_{\rm grav} + \Delta F_{\rm cori})$.
	Here, the Coriolis force perturbation $\Delta F_{\rm cori}$ is calculated as $-2 \Omega \times (v-v_{\rm cir})$.
}
\label{fig:VelocityProfile}
\end{center}
\end{figure*}

The swing amplification mechanism can explain certain aspects of the evolution of spiral arms,
i.e., the amplification (excitation) of density enhancement.
However, the destruction process of the non-steady spirals, as seen in 
Figure \ref{fig:snapshotModelMSdamping} \citep[see also][]{Fujii+2011,Wada+2011},
cannot be understood only by the swing amplification mechanism.
The top right panels in Figure \ref{fig:VelocityProfile} show 
the time evolution of the relative velocities ($\Delta v_\eta$ and $\Delta v_\xi$)
in the co-rotating frame of the spiral arm.
Upon considering the coordinates $(\eta, \xi)$,  i.e., 
the $\eta$- and $\xi$-axes are perpendicular and parallel to the spiral arm, respectively, 
the velocity component along each is given by
\begin{eqnarray}
 && \Delta v_{\eta} \equiv  v_{R} \cos{i} + (v_{\phi} - v_{\rm cir})\sin{i},\\
 && \Delta v_{\xi}   \equiv -v_{R} \sin{i} + (v_{\phi} - v_{\rm cir})\cos{i},
\end{eqnarray}
where $i$ is the pitch angle of the spiral arm, and $v_{\rm cir}$ is the 
circular velocity determined by the azimuthally averaged gravitational field.
For this definition, inflow to the arm corresponds to $\Delta v_{\eta} > 0$ for $\eta < 0$, 
and $\Delta v_{\eta} < 0$ for $\eta > 0$.
In the amplification phase ($T_{\rm rot} \lesssim 12.2$),  we see a clear inflow motion
to the spiral arm along both sides of the chosen strip (solid lines).
In the initial stages of the damping phase ($T_{\rm rot} \simeq 12.3$), 
the streaming velocity around the spiral arm gradually transits from inflow to outflow.
At the end of the damping phase ($T_{\rm rot} \simeq 12.5$), 
it is clear that the stars in the spiral arm move away from the arm along both its sides 
(dot-dashed lines).
During the dynamical evolution, the parallel component of the velocity, $\Delta v_{\xi}$, 
does not change its sign or show a decrease its magnitude. 
The left panels of Figure \ref{fig:VelocityProfile} show a schematic view of the time evolution 
of the non-circular velocity field associated with the non-steady spiral arm.

The middle right panels in Figure \ref{fig:VelocityProfile} show the time evolution of 
the non-axisymmetric gravitational forces (i.e., spiral perturbation) 
involved in the damping phase.
The component of the non-axisymmetric gravitational force perpendicular to 
the spiral arm ($\Delta F_{\rm grav,\eta}$), 
which is stronger than the parallel component ($\Delta F_{\rm grav,\xi}$), 
is always directed towards the arm 
($\Delta F_{\rm grav,\eta} > 0$ for $\eta < 0$, and $\Delta F_{\rm grav,\eta} < 0$ for $\eta > 0$)
during the dynamical evolution.
The bottom right panels in Figure \ref{fig:VelocityProfile} show the time evolution of 
the net force (i.e., non-axisymmetric gravitational force plus Coriolis force perturbation). 
The component of the net force perpendicular to the spiral arm ($\Delta F_{\rm grav+cori,\eta}$)
evolves in the same manner as $\Delta F_{\rm grav,\eta}$ 
because there is a strong density gradient along the $\eta$-direction.
The parallel component $\Delta F_{\rm grav,\xi}$ is almost zero during the growth
and damping phases and does not change. This is because the density
gradient along the $\xi$-direction is small and almost unchanged.
However, the parallel component of the net force $\Delta F_{\rm grav+cori,\xi}$ changes its
$\eta$-dependence from the growing phase ($T_{\rm rot} = 12.1$) 
to the damping phase ($T_{\rm rot} > 12.3$)
according to the change of the sign of the perpendicular velocity ($\Delta v_{\eta}$),
since the Coriolis force works perpendicular to the direction of the velocity.

This indicates that the Coriolis force exerted on the stars in the damping phase
exceeds the non-axisymmetric gravitational force due to the spiral perturbation.
This causes stars to ``escape'' from the spiral perturbation, and eventually 
the spiral arm itself begins to thin out and fade.

The above argument suggests that the non-steady nature of stellar spirals is originated in 
the evolution of the orbits of the stars in the spirals. 
The phenomenon of swing amplification \citep{Toomre1981} is a part of this 
non-linear coupling between particles and waves, but it does not 
describe all observed phenomena.
In the next section, we explore the orbital evolution of stars
associated with the  spiral arm in detail.

\section{Orbital Evolution of Stars Around Spiral Arms}
\label{sec:StarAroundSpiralArm}

\begin{figure*}[p]
\begin{center}
\includegraphics[width=.90\textwidth]{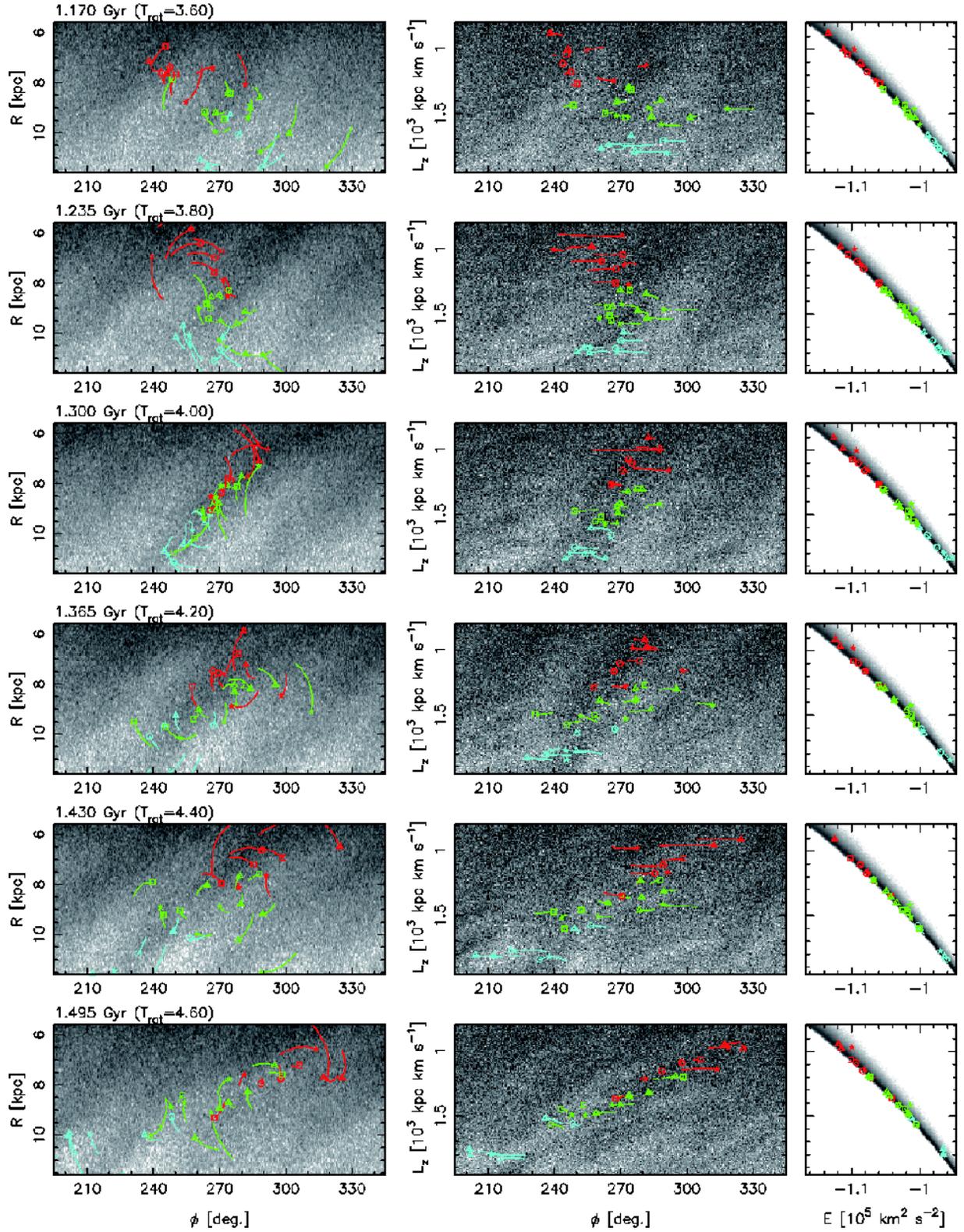}
\caption{
	Orbital evolution of stars in spiral arm. 
	The stars associate around the spiral arm within a distance of $\pm 0.5$ kpc
	at $T_{\rm rot} = 4.0$.
	Left columns: orbits on $\phi-R$ plane. 
	Middle columns:  orbits on $\phi-L_{\rm z}$ plane.
	Right columns:  orbits on $E-L_{\rm z}$ plane.
	The colors denote the angular momentum at the time instants 
	when the stars associated with the spiral arm.
	See the attached video.
}
\label{fig: orbit_linear}
\end{center}
\end{figure*}

Figure \ref{fig: orbit_linear} shows the evolution of stars along the $\phi-R$ plane in
the early phase of spiral development. 
At this stage of the simulation, 
we selected 15 particles associated with 
one of the three weak spiral arms that had evolved at $T_{\rm rot} = 4.0$.
The figure shows that stars with epicycle motion are captured 
by the density enhancement ($T_{\rm rot} = 3.6-4.0$),
and further, these stars are dissociated from the original arm ($T_{\rm rot} = 4.2-4.6$).
This behavior is similar to that of the ``density wave'' from a certain viewpoint, 
but the spiral arms are short-lived and co-rotating (left panel of Figure \ref{fig:PatternSpeed}).
Thus, even if the arms appeared in the early linear phase, they are not completely explained
by the picture of stationary density waves.
The middle column of images in Figure \ref{fig: orbit_linear} shows the plots of 
the azimuth angle ($\phi$) versus the angular momentum $L_{z}$ curve instead of the $\phi-R$ plot.
It is clear that all the stars along the $\phi-L_{z}$ plane oscillate horizontally
when the angular momentum of each star is conserved.
The images in the right column of Figure \ref{fig: orbit_linear} show the so-called Lindblad diagram, 
where the angular momentum $L_{z}$ of each star is plotted against its total energy $E$.
Most of the stars  show no significant movement from their original position, 
thereby suggesting that their energy is also conserved.

\begin{figure*}
\begin{center}
\includegraphics[width=.90\textwidth]{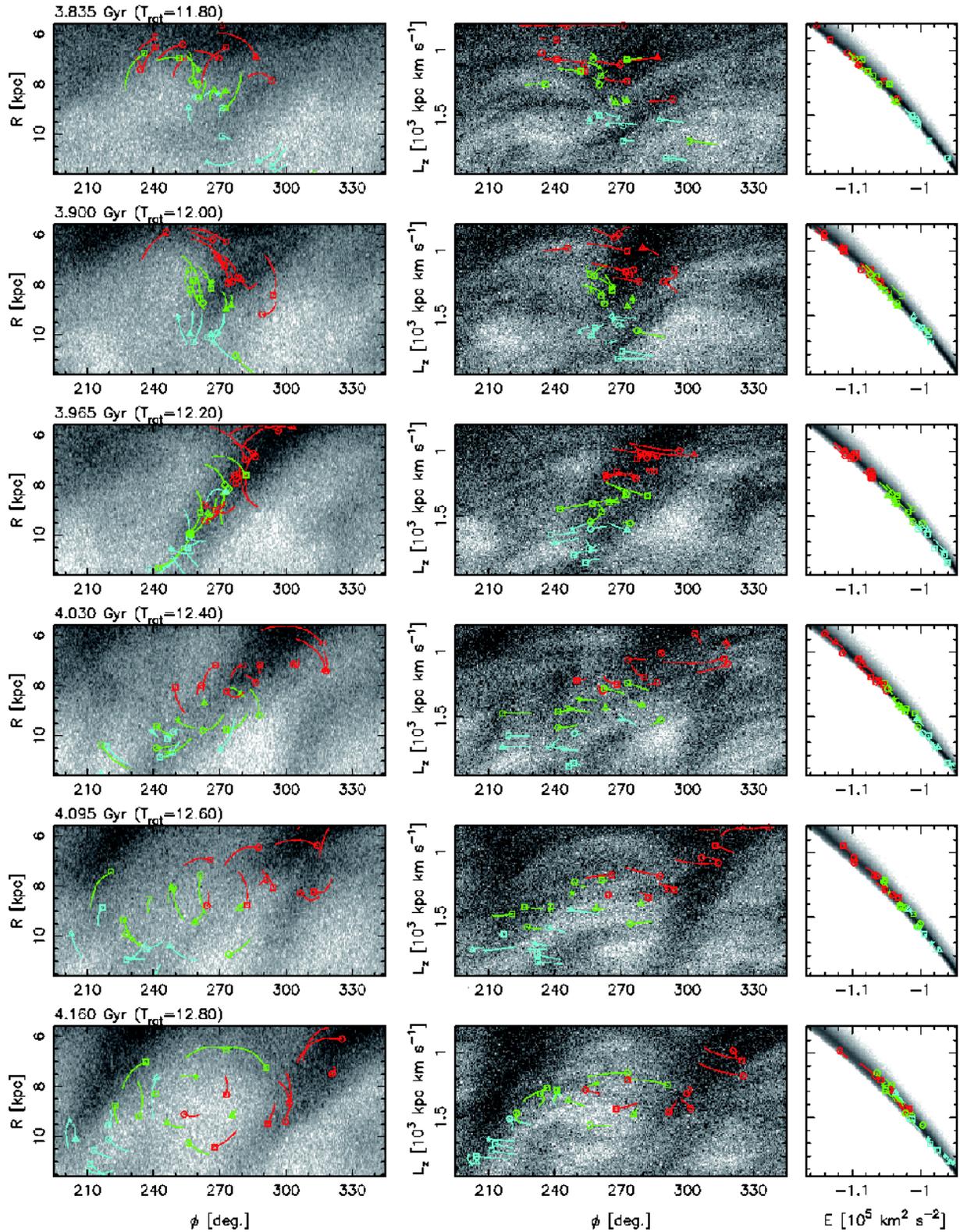}
\caption{
	Orbital evolution of stars in spiral arm during non-linear phase.
	The stars associate with the spiral arm at $T_{\rm rot} = 12.2$.
	See the attached video.
	}
\label{fig:EvolutionSpiral}
\end{center}
\end{figure*}

On the other hand, the behavior of the stars is very different in the non-linear phase, 
where the spiral arms are well-developed and non-steady (refer to previous sections).
Figure \ref{fig:EvolutionSpiral} shows images identical to those in Figure \ref{fig: orbit_linear};
however, in this set of images, the stars are in motion due to change in
their angular momenta in the non-linear phase of orbital evolution (see also the supplementary video).
When the stars are captured by the density enhancement ($T_{\rm rot} \simeq 11.8-12.0$), 
they radially migrate along the spiral arms.
The stars approaching from behind the spiral arm (i.e., inner radius) tend to 
attain increased angular momenta via acceleration along the spiral arm, 
whereby they move to the disk's outer radius.
In contrast, the stars approaching ahead of the spiral arm (i.e., outer radius) tend to
lose their angular momenta via deceleration along the spiral arm, 
and they move to the disk's inner radius.
Along the $\phi-L_{z}$ plane, the stars oscillate both horizontally as well as vertically. 
Moreover, the guiding centers of the oscillations do not remain constant at the same 
value of $L_{z}$. 
This is essentially different from the epicycle motion in which $L_{z}$ is conserved.

The Lindblad diagram in Figure \ref{fig:EvolutionSpiral} shows that
the stars oscillate along the curve of circular motion
by undergoing change in terms of both angular momentum and energy.
The oscillating stars successively undergo aggregation and disaggregation along the curve, 
thereby leading to the formation of structures
referred to as ``swarms of stars'' along the  $\phi-L_{z}$ and $R-\phi$ planes.
The non-steady nature of the spiral arms originates in the dynamical 
interaction between these ``swarming'' stars with a {\it non-linear} epicycle motion and 
the high-density regions, i.e., the spiral arms moving with the galactic rotation.
This is similar to the wave-particle interaction described previously; 
however, in this case, the high-density regions are not ``waves''.
Since the non-steady spirals move at the rate of the local galactic rotational speed (Section 3),
in contrast to a spiral perturbation with a single pattern speed,
co-rotating points are found everywhere on the spiral arms. 
Therefore, the motion of stars along the $E-L_{z}$ plane can be naturally understood
as due to their scattering around co-rotating spirals 
\footnote{
Stars around the co-rotation point 
change their angular momenta without increasing their random energy
\citep{LyndenBellKalnajs1972}.
}.
This behavior is similar to reported in previous studies
\citep[e.g.,][]{SellwoodBinney2002,Grand+2012,Bird+2011,Grand+2012b}.
Furthermore, we observe that the $E-L_{z}$ curve of each star changes over a large range of radii.
This is entirely different from what is expected in stationary density waves, 
where these changes are limited to the Lindblad resonances \citep{LyndenBellKalnajs1972}.
Further, it is noteworthy that the structures formed self-induced 
in the angular momentum space; this is a property similar to that of the ``groove'' mode 
hypothesized by \citet{SellwoodLin1989}.
This point is beyond the scope of this paper. We will investigate this elsewhere.

\begin{figure*}
\begin{center}
\includegraphics[width=.8\textwidth]{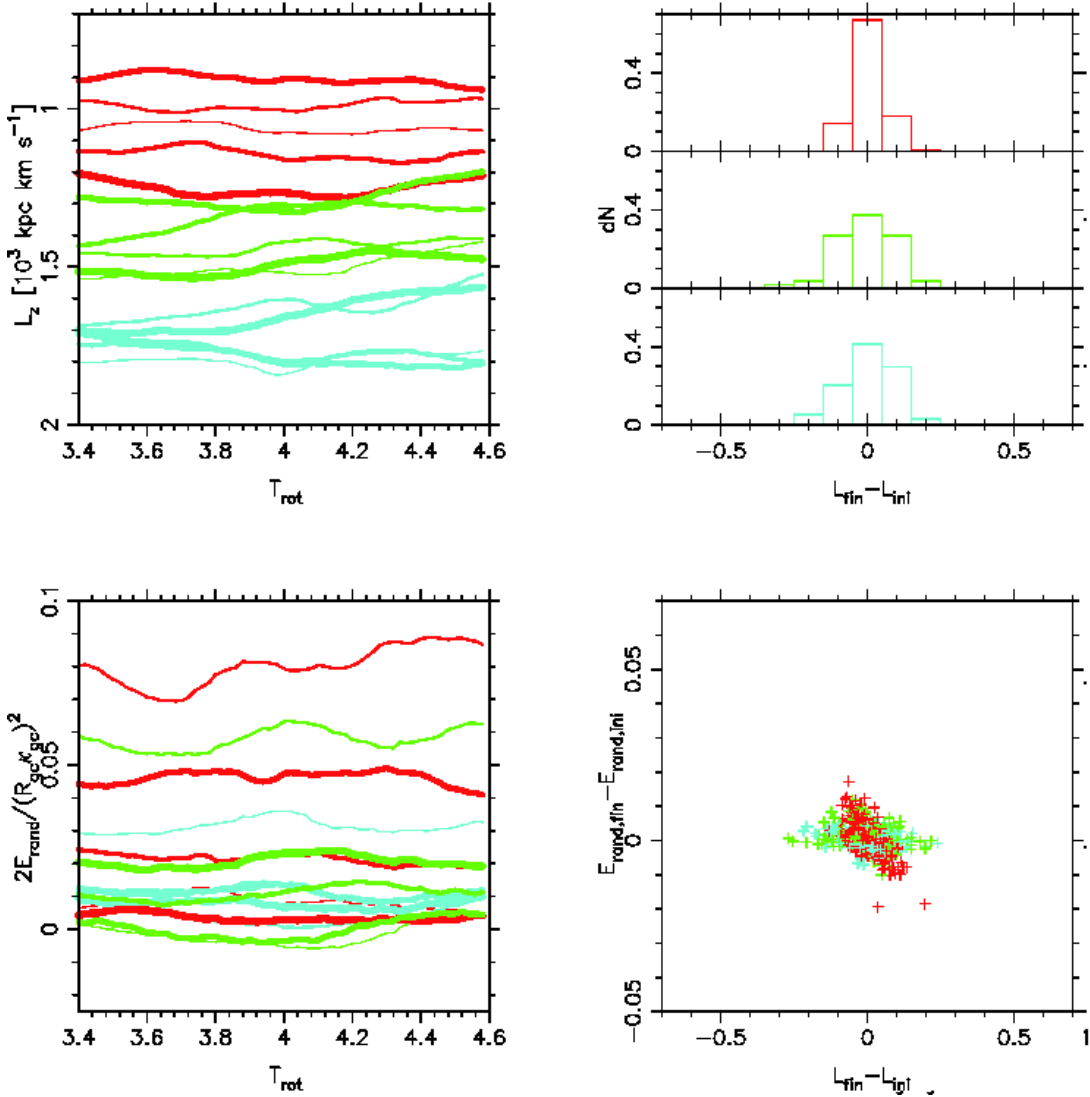}
\caption{
	Time evolution of orbital characters of stars around spiral arm during 
	early evolutional stage shown in Figure \ref{fig: orbit_linear}. 
	Left: Time evolution of the orbital angular momentum (top) and the random energy (bottom) of the stars. 
	The plotted stars are the same as those in Figure \ref{fig: orbit_linear}.
	The random energy of the stars is non-dimensional 
	using the radius of the guiding center $R_{\rm gc}$ and epicyclic frequency $\kappa_{\rm gc}$,
	which is proportional to the square of the orbital ellipticity $e$.
	Each star is indicated by a different line thickness.
	The colors correspond to those of Figure \ref{fig: orbit_linear}.
	Right: Angular momentum change, $L_{\rm fin}-L_{\rm ini}$, 
	distributions of stars (top). Here, $\Delta T_{\rm rot} = 0.8$ is adopted 
	to measure the angular momentum change and
	the vertical axis indicates the normalized number of particles.
	The random energy changes $E_{\rm rand,fin}-E_{\rm rand,ini}$, 
	in the stars over $\Delta T_{\rm rot} = 0.8$ as a function of 
	their angular momentum changes (bottom). 
	The colors correspond to those in the left panels. 
	}
\label{fig:L-t04}
\end{center}
\end{figure*}

\begin{figure*}
\begin{center}
\includegraphics[width=.8\textwidth]{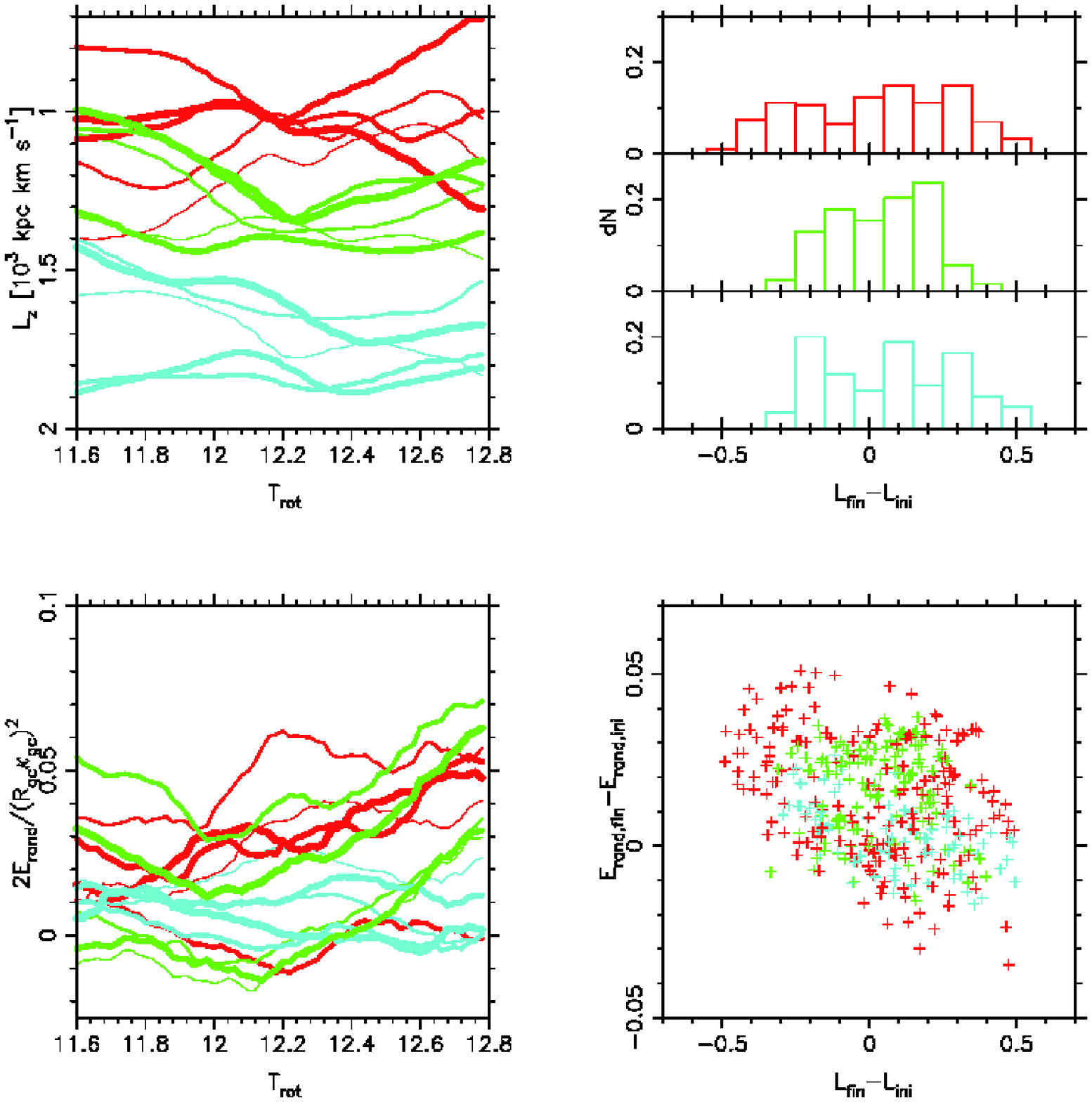}
\caption{
	Time evolution of orbital characteristics of stars for phase in which the spiral arms
	are non-linearly developed (shown in Figure \ref{fig:EvolutionSpiral}).
	The plotted stars are the same as those in Figure \ref{fig:EvolutionSpiral}.
}
\label{fig:L-t12}
\end{center}
\end{figure*}

The time evolutions of the angular momentum of stars associated 
with a spiral arm at $T_{\rm rot} \simeq 4.0$ and $T_{\rm rot} \simeq 12.2$
are plotted in the left top panels of Figures \ref{fig:L-t04} and \ref{fig:L-t12}, respectively.
In the early phase ($T_{\rm rot} \simeq 4.0$),  
both the angular momentum and random energy do not change
by more than 10\% during one rotational period (top right panel of Figure \ref{fig:L-t04}).
On the other hand, in the non-linear phase ($T_{\rm rot} \simeq 12.2$), 
the fraction of the angular momentum changes by $\sim$ 50\% (top right panel of Figure \ref{fig:L-t12}).
It is clear that the angular momentum of the stars changes
significantly due to their scattering by a well-developed spiral at $T_{\rm rot} = 12.2$.
This corresponds to the hypothesis that the guiding center of the epicycle motion of each star 
undergoes radial motion.

The change in the normalized random energy shown in 
the left bottom panels of Figures \ref{fig:L-t04} and \ref{fig:L-t12} indicates 
that certain stars with relatively small initial random energies 
experience a large energy change $\Delta E_{\rm rand}$ after interaction with a spiral arm.
Here, the random energy is calculated as the difference between the total energy and 
the circular energy, i.e., $E_{\rm rand} = E - E_{\rm cir}(R_{\rm gc})$.
It is noteworthy that the random energy change is {\it not} always positive;
a significant fraction of stars {\it lose} their random energy.
This is because the perturbation from the spiral arm shifts the guiding center
of the stars' epicycle motion without increase in orbital eccentricity.
It is to be noted that this argument is rigorously correct; in fact, Figure \ref{fig:L-t12} (bottom left panel) 
shows the changes in the random energy of the stars as a function of change in the angular momentum.
We can see a weak trend: the outer migrators (i.e., $L_{\rm fin}-L_{\rm ini}>0$) undergo a decrease 
in their random energy ($E_{\rm rand,fin}-E_{\rm rand,ini}<0$), and vice versa.
A similar effect of the radial migration of stars around spiral arms upon disk heating has been noticed 
in recent numerical simulations \citep{Grand+2012,Roskar+2011,Grand+2012b,Minchev+2012}.

In summary, the gravitational interaction between the stars in the spiral arm and 
the spiral density enhancement changes the angular momentum and random energy of the stars, 
and this process in turn changes the structure of the spirals.
During this process, the random energy of individual stars in the system 
does {\it not} increase monotonically. 
In other words, local interactions between the non-steady arms and stars 
increase or decrease the total energy of individual stars locally;
however, the energy remains around its value for circular motion with the occurrence of a small dispersion.
This is because the interaction causes the migration of the guiding centers of the stars 
without increasing their eccentricity or random energy.
This ``dynamical cooling'' mechanism is essential to
prevent heating of the stellar disk and erasure of the spiral arms, and 
the mechanism produces ``swarms'' of stars moving between non-steady spirals.
The non-linear epicycle motion of the stars and their non-linear coupling with the density perturbation
is the fundamental physics of the recurrently formed, non-steady spiral arms in a stellar disk.

\section{Discussion}
\label{sec:Discussion}

\begin{figure}
\begin{center}
\includegraphics[width=.4\textwidth]{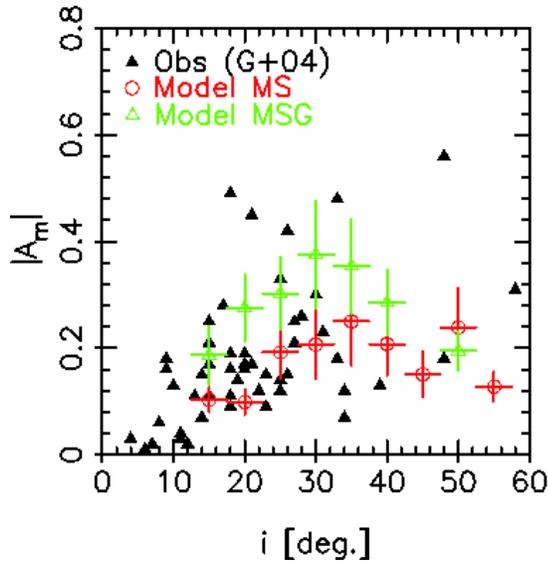}
\caption{
	Pitch angle-amplitude correlation.
	Morphological properties (amplitudes $A_{\rm m}$ and pitch angles $i$) 
	of stellar spiral arms are derived using a 2D Fourier fitting to 
	the range of radii given by $R_{\rm sd} < R < 3R_{\rm sd}$. 
	Here, $|A_{m}|$ values for the $m=4$ mode in the simulations are shown.
	Model MSG is nearly identical to Model MS; 
	however, model MSG also includes the gaseous component 
	with a mass fraction that is 10\% of the stellar disk mass \citep[refer to][]{Wada+2011}.
	The filled triangles represent the observational data obtained from the study by \citet{Grosbol+2004}.
}
\label{fig:spiralproperties}
\end{center}
\end{figure}

Figure \ref{fig:spiralproperties} compares the morphological properties (the pitch angle $i$ and
the amplitude $|A_{m}|$) of the stellar spiral arms between simulations and observations.
For the comparison, we used a 2D fast Fourier transform (2D FFT) analysis with logarithmic spirals
\citep[for details, see for e.g.,][]{PuerariDottori1992}.
The plotted points in green and red symbols represent the time evolution of the spiral arms
for the two models that are essentially identical (see footnote).
We found that the simulated spiral arms exhibit a maximum amplitude around $i \sim 30^{\circ}$. 
Further, we found that the spiral arms tend to be weaker in the pure stellar disk model (model MS) than
the model with the ISM (model MSG)
\footnote{
Model MSG is the same model as that presented by \citet{Wada+2011}.
This model is mostly identical to model MS; however, in model MSG, 
the ISM, star formation, and supernova feedback are also taken into account.
The hydrodynamics of the model is solved by the smoothed particle hydrodynamics (SPH) method. 
The initial gas mass fraction is 10 \% of the stellar disk mass, and 
the initial density profile follows an exponential profile with a scale-length 
twice that of the stellar disk. Refer to \citet{Wada+2011} for details.
}. 
This distribution which has a peak at around $i \sim 20-30^\circ$ is qualitatively consistent with 
the the prediction of the swing amplification mechanism (see section 4.1).
We compared the simulated distribution with the observed one.
The dependence of $|A_{m}|$ on the pitch angle 
in the models with/without the ISM is consistent with observations at least for 
the range of values corresponding to $i < 30^{\circ}$.
This comparison suggests that the self-induced spiral arms in 
differentially rotating disks and their time evolution can fairly consistently 
account for the morphological diversity seen in nearby spiral galaxies.

However, this comparison between the simulated and observed distributions requires further examination.
Firstly, the determination of the pitch angle is somewhat uncertain. 
The analyzed pitch angles differ between studies for the same galaxy.
For example, the pitch angles of NGC 3054 are given by $33^\circ$, $43^\circ$, and $12^\circ$ in
 \citet{Grosbol+2004}, \citet{Seigar+2006}, and \citet{Davis+2012}, respectively.
Secondly, there is a lack of statistical studies on the relation between 
the pitch angle and the amplitude of the stellar spiral arms.
Although some statistical studies have used 2D FFT to analyze the pitch angle of the spiral arms,
they do not include examination of the amplitude or contrast of the spiral arms 
\citep[e.g.,][]{Seigar+2005,Seigar+2006,Davis+2012}.
Thus, further progress in understanding the spiral dynamics requires 
additional statistical and robust observation data of the morphological properties 
(both the pitch angle and amplitude) of the stellar spiral arms.
This relation between the pitch angle and amplitude will form one of the tests for spiral genesis theories
along with the pitch angle-shear rate (or Hubble-type) relation 
\citep{Roberts+1975,SeigarJames1998,Hozumi2003,Grand+2012c}, 
existence of systematic angular offset between the young stellar component
and the stellar spiral arm \citep{Fujimoto1968,Roberts1969,Egusa+2009,Foyle+2011,Ferreras+2012}, 
and the radial dependence of the pattern speed \citep{Meidt+2008,Meidt+2009}.

Finally, we comment on grand-design spirals.
In a series of studies \citep[][]{Fujii+2011,Wada+2011} including this one, 
we have examined the stellar dynamics of non-steady stellar spirals, 
as well as the interactions between them and the ISM. 
However, these studies have focused more on multi-armed spirals than grand-design spirals 
(i.e.,  $m=2$ spirals) whose fraction is more than $\sim 50\%$ 
in nearby spiral galaxies \citep{Grosbol+2004,Kendall+2011}.
It has been observed that grand-design spirals are associated with bars or companions 
\citep{KormendyNorman1979,SeigarJames1998,Salo+2010,Kendall+2011}, 
and this observation is consistent with the results of many numerical simulations 
of bar-driven spirals \citep[e.g.,][]{SellwoodSparke1988,Bottema2003}
and tidally-driven spirals \citep[e.g.,][]{Oh+2008,Dobbs+2010,Struck+2011}.
We will focus on bar-driven spirals and tidally-induced spirals in forthcoming studies.

\section{Summary}

The $N$-body simulations of an isolated disk galaxy show the formation of self-induced, 
non-steady multi-arm spirals that follow the differential galactic rotation.
We found that the swing amplification mechanism causes the development of the spirals. 
When a spiral undergoes the damping phase, the Coriolis force dominates the gravitational
perturbation exerted by the spiral, and  as a result, stars escape from the spirals, and join a
new spiral at a different position.
This process uniformly for a given spiral, thereby resulting in the formation of bifurcating
and merging spiral arms; therefore, the dominant spiral modes always show change 
in their radii over time. 
We confirmed that this phenomenon originates due to the changing orbital properties of stars. 
The angular momentum and energy of each star undergo changes 
due to the star's interaction with the spiral arms.
As a result, the epicycling stars radially migrate; in other words, their guiding centers also undergo motion.
Interestingly, the movement of groups of stars with similar orbital properties 
causes the appearance of ``swarming''. 
In the non-linear phase of development of spiral instability, the swarming stars 
cause complicated morphological changes in the spiral arms.

During this process, the random energy of 
individual stars (or orbital eccentricity) does {\it not} increase monotonically.
In fact, a significant fraction of stars even lose their random energy.
This ``dynamical cooling'' due to the mechanism like the wave-particle interaction 
can explain why the short-lived spiral arms are self-induced over several 
rotational periods despite the absence of dissipative component in the disk \citep{Fujii+2011}.

In the above process, it is essential that spiral arms mostly follow the galactic rotation at any radius;
in other words, the ``co-rotating points'' are required to be ubiquitous in the differentially rotating galactic disk.
The conclusions of other previous studies also indicate the possibility that
the co-rotation resonance affects stellar motions more in terms of 
radial migration than heating up of the disk 
\citep{LyndenBellKalnajs1972,SellwoodBinney2002,Grand+2012,Grand+2012b,Roskar+2011,Minchev+2012}.

We conclude that the non-linear epicycle motions and self-gravity 
in the differential rotation of stellar disks are essential for 
the recurrent amplification and destruction processes of the spiral arms.
In other words, the issue of  the so-called ``winding dilemma''  is no longer a problem 
at least in multi-armed spiral galaxies.

\acknowledgments
The authors are grateful to the anonymous referee for his/her valuable comments.
We would like to thank Michiko S. Fujii, Junichiro Makino, Shunsuke Hozumi, and Daisuke Kawata.
Numerical simulations were performed using the Cray XT-4 at the Center for 
Computational Astrophysics (CfCA), National Astronomical Observatory of Japan. 
This research was supported by the HPCI Strategic Program Field 5 
``The Origin of Matter and the Universe.'' 
This work was supported in part by Grant-in-Aid for Scientific Research (C)23540267.

\bibliographystyle{apj}
\bibliography{ms}

\end{document}